\documentclass{ws-ijmpd}

\begin{document}

\markboth{E. M. de  Gouveia Dal Pino, G. Kowal, L. H. S. Kadowaki, P. Piovezan, A. Lazarian}
{Magnetic Field Effects near the launching region of Astrophysical Jets}

%
\catchline{}{}{}{}{}
%

\title{Magnetic Field Effects around the launching region of Astrophysical Jets}

\author{E. M. de  Gouveia Dal Pino, G. Kowal, L. H. S. Kadowaki, P. Piovezan}

\address{Instituto de Astronomia, Geof\'isica e Ci\^encias Meteorol\'ogicas (IAG), Universidade de S\~ao Paulo, Rua do Mat\~ao 1226, Cidade Universit\'aria,\\
S\~ao Paulo, SP 05508-900, Brazil\\
dalpino@astro.iag.usp.br}

\author{A. Lazarian}

\address{Astronomy Departament, University of Wisconsin, Madison, USA}

\maketitle

\begin{history}
\received{Day Month Year}
\revised{Day Month Year}
\comby{Managing Editor}
\end{history}

\begin{abstract}
One of the fundamental properties of astrophysical magnetic fields is their ability to change topology through reconnection and in doing so, to release magnetic energy, sometimes violently. In this work, we review recent results on  the role of magnetic reconnection and associated heating and particle acceleration in jet/accretion disk systems, namely young stellar objects (YSOs), microquasars, and active galactic nuclei (AGNs).
\end{abstract}


\section{Introduction}	

Supersonic jets are observed in several astrophysical systems, such as active galactic nuclei (AGNs), neutron star and black hole X-ray binaries, low mass young stellar objects (YSOs), and are probably also associated with gamma ray bursts.

The most accepted mechanism for jet production is based on the magneto-centrifugal acceleration out of a rotating magnetized accretion disk that surrounds the central source. Firstly proposed by Blandford \& Payne (1982), this basic scenario for jet launching and variants of it have been object of extensive analytical and numerical investigation (see e.g., McKinney \& Blandford 2009 and de Gouveia Dal Pino 2005 for reviews). However, though considerable progress has been achieved in the comprehension about the possible origin of the magnetic fields that must permeate the accretion disk and the mechanism
of angular momentum transport that allows the accretion to occur through magnetorotational turbulence (Balbus \& Hawley 1998),  there are still fundamental questions to be solved. For instance, we do not understand yet the nature of the coupling between the central source magnetosphere and the disk field lines, nor  the origin of the quasi-periodic ejections that are often associated to these jets.

In 2005, de Gouveia Dal Pino \& Lazarian proposed a mechanism that could be responsible for non-steady ejections of relativistic plasma
 in microquasars.
Their scenario is related to violent reconnection episodes between the magnetic field lines of the inner disk region and those that are anchored into the black hole.
In recent work, we have extended this mechanism to the AGN jets and to thermal YSO jets (de Gouveia Dal Pino et al. 2010).
In this work we review the basic physics behind the magnetic reconnection process and highlight the current limitations of the model when applied to the launching base of jets in inner accretion disk coronal regions.

\section{Fast magnetic reconnection}

A magnetic field embedded in a perfectly conducting fluid preserves its topology for all times. Although ionized astrophysical objects, such as stars and galactic disks, are almost perfectly conducting fluids, they show indications of changes in topology through magnetic reconnection within dynamical time scales. Reconnection can be observed directly in the solar corona, but can also be inferred from the existence of large scale dynamo activity inside stellar interiors. Solar flares and gamma ray busts are usually associated with magnetic reconnection.

In previous work, Lazarian \& Vishniac (1999) showed how reconnection can be rapid in plasmas with very small collision rates due to turbulence. More recently,  three dimensional simulations (Kowal et al. 2009) have confirmed  the predictions of this earlier work and evidenced that reconnection in a turbulent fluid occurs at a speed comparable to the rms velocity of the turbulence, regardless of the value of the resistivity. In particular, this is true for turbulent pressures which are much weaker than the magnetic field pressure so that the magnetic field lines are only slightly bent by the turbulence. These results are consistent with the proposal by Lazarian \& Vishniac (1999) that reconnection is controlled by the stochastic diffusion of the magnetic field lines, which produces a broad outflow of plasma from the reconnection zone. This result implies that reconnection in a turbulent fluid typically takes place in approximately a single eddy turnover time. This has important implications for dynamo activity and particle acceleration in different astrophysical environments. We will discuss this point further in session 4 below. We note that, in contrast, the reconnection in 2D configurations in the presence of turbulence depends on resistivity, i.e. is slow (Kowal et al. 2009).

\section{Jet launching region around YSOs and the role of magnetic reconnection}

YSOs differ in many aspects from microquasars and AGNs. For instance, they produce thermal rather than relativistic jets and exhibit emission lines from which their physical properties (such as density and temperature) are inferred.
Because of their proximity and frequency in star formation regions, YSO jets are usually regarded as excellent laboratories for jet investigation. Recent years have presented the first attempts to observe any evidence of rotation of these jets at their launching region as a probe of the magnetic centrifugal models.  Bacciotti et al. (2002), for instance, measured  Doppler-shift asymmetries in the transverse velocity of $10-25$ km/s in micro-jets at few 10 AU from the origin that were interpreted as evidence of jet rotation. However, 3D numerical simulations of $non-rotating$ precessing jets by Cerqueira et al. (2006) have produced the same Doppler shift patterns at the transverse velocity. Also, Soker (2009) argued that a similar pattern could arise from interaction of the jet with circumstellar gas and besides, the expected rotation velocities at those distances above the accretion disk should be much smaller. This means that we will have to wait for the coming new generation of more sensitive instruments (e.g., ALMA and LLAMA) in order to probe jet rotation at the launching region.

Other important clues for the jet launching mechanism rely on the nature of the interaction between the central source and the accretion disk.
There is observational evidence that accretion onto young stars is dynamic on several timescales and governed by changes in the magnetic field configuration (see Alencar, Johns-Krull \& Basri 2001 and Alencar 2007). Magnetic fields of 1 to 3 kG have been measured at the surface of T Tauri stars and they seem to be strong enough to disrupt the inner disk and allow the channeling of magnetospheric accretion in funnel flows (see Figure 1). It turns out that magnetospheric accretion through the funnel flows explains main observed characteristics of (T Tauri) YSOs (Camenzind 1990).
At the same time, numerical simulations predict evolving funnel flows due to the interaction between the stellar magnetosphere and the inner disk region.
The differential rotation between the star and the inner disk regions lead  to field lines opening and reconnection which eventually restores the initial magnetospheric configuration (e.g., Goodson \& Winglee 1999 and Romanova et al. 2004; see also Alencar 2007 for a review).
Magnetospheric reconnection cycles are expected by most numerical models to develop in a few Keplerian periods at the inner disk, and be accompanied by time dependent accretion onto the star and by episodic outflow events as reconnection takes place.

On the other hand, YSOs  may exhibit intense magnetic activity that results in a strong and variable x-ray emission (Bouvier et al. 2007). Observed x-ray flares  are often attributed to magnetic activity at the stellar corona (Feigelson \& Montmerle 1999). However, some COUP (Chandra Orion Ultra-deep Project) sources have revealed strong flares that were related to peculiar gigantic magnetic loops linking the magnetosphere of the central star with the inner region of the accretion disk. It has been argued that this x-ray emission could be due to magnetic reconnection in these gigantic loops (Favata et al. 2005). de Gouveia Dal Pino et al. (2010) have recently examined this issue in more detail  investigating the role of magnetic reconnection events in the inner disk/corona of YSOs to explain the gigantic x-ray flares and check if magnetic reconnection may be related to the thermal jets of these sources.

In order to estimate the amount of magnetic power that may be released by violent reconnection in the inner disk-star region they considered a most simple geometrical configuration to the problem as in Figure 1.  To evaluate the disk quantities in this region, they adopted the standard model (see Shakura \& Syunyaev 1973 for more details), assuming a pressured dominated disk. For the coronal region above the disk, they have considered mean values of density and temperature inferred from observations (Favata et al. 2005). The stellar magnetic field was assumed to be quasi dipolar. The equilibrium between the stellar magnetic pressure and the disk ram pressure in the inner disk region, gives the radius at which the disk is truncated. With these assumptions, the maximum possible accretion rate is reached when the disk touches the stellar surface and it can be estimated by
$\dot{M}_{max} = 1.12\times 10^{22} \left( \mu B_{*}\right)_{5000}^2 R_2^{5/2} M_1^{-1/2}$ g/s,
where $\left(\mu B_* \right) = 5000 \left(\mu B_* \right)_{5000}$ G, $R_* = 2
R_{\odot} R_2$, and $M_* = 1M_{\odot} M_1$ is the stellar mass.
Associated to this maximum accretion rate there is a maximum magnetic power
released by magnetic reconnection events (de Gouveia Dal Pino et al. 2010):


\begin{equation}
\dot{W}_{B,max}=1.4 \times 10^{31} \left( \mu B_* \right)_{5000}^{-16/7} M_1^{11/7} \dot{M}_{max}^{22/7} R_{2}^{-48/7}\times ~ n_{11}^{-3/2} T_8^{-1/2} ~~ erg/s
\end{equation}
 where $n_c = 10^{11} n_{11}$ $cm^{-3}$ is the coronal density, and $T_c = 10^8
T_8$ K is the coronal temperature, both estimated from observations (Favata et al. 2005).

A comparison between the predicted maximum magnetic power released under these circumstances in violent reconnection events and the observed x-ray luminosities for a sample of COUP sources (Favata et al. 2005) have shown that for most sources, the magnetic power is of the order of (or greater than) the observed luminosities if the maximum accretion rate is $(0.5-1)\times 10^{-4}$ $M_{\odot}/yr$ (for 10 sources) or
$(2.5 - 5) \times 10^{-4}$ (for 5 sources) (see Fig. 4 of de Gouveia Dal Pino et al. 2010).
Such accretion rates of  about $100-1000$ times greater than the mean typical rates expected for evolved YSOs, would be required only for
very short time intervals during violent magnetic reconnection events.
This implies that the detection of other spectral signatures of such
higher accretion rates would be probably very difficult.
The energy released by violent magnetic reconnection processes could also help to heat the gas at the base of the jets (Ray et al. 2007)\cite{}. A rough estimate indicates that the large amount of magnetic power released by magnetic reconnection events can be thermally conducted up to distances $\sim 10$ AU in a timescale $\tau _{cond}
\sim 10^8 n_{10} T_{8}^{-5/2} l_{10}^2$ s that is comparable, e.g. to the dynamical
timescale of the DGTau jet (e.g., Bacciotti et al. 2002 and Cerqueira \& de Gouveia dal Pino 2004).

\begin{figure}
 \begin{center}
  \includegraphics[width=0.45\textwidth]{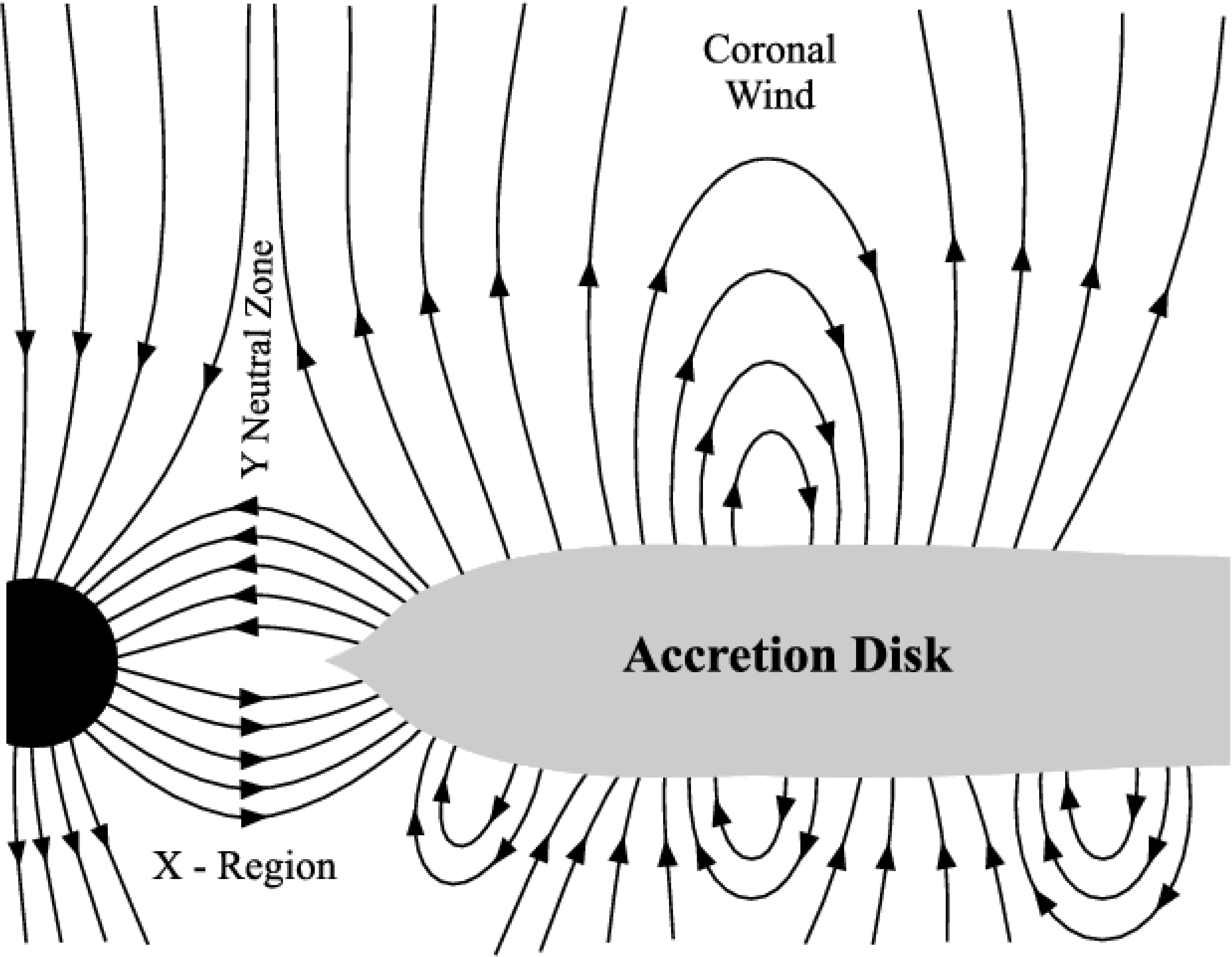}
  \caption{Schematic drawing of the magnetic field geometry in the inner disk-source  region at $R_X$. From de Gouveia Dal Pino \& Lazarian (2005).}
  \label{fig:fig1}
 \end{center}
\end{figure}

\section{Magnetic reconnection effects in the launching region of relativistic jets}

Relativistic jets from stellar-mass black holes of binary stars
emitting X-rays, also called microquasar jets (or
BHXRB jets), are scaled-down versions of AGN (or quasar) jets,
typically extending for $\sim$ 1pc and probably powered by the accreting,
spinning black hole. Despite the enormous difference in scale, both
classes share several similarities in their physical properties.

According to their
X-ray emission (2-20 keV), all classes of BHXRBs show basically two major states: a
quiescent and an outburst state. The former is characterized by low
x-ray luminosities and hard non-thermal spectra.
On the other hand, the outburst state corresponds to intense
activity and emission and can be sub-classified in three main active
states, the Thermal State (TS), the
Hard State (HS) and the Steep Power Law State (SPLS) (Remillard \& McClintock 2006). These states
are usually explained as changes in the structure of the accretion
flow. During the TS, for example, the soft x-ray thermal emission
is believed to come from the inner region of the thin accretion disk that extends
till the last stable orbits around the black hole. On the other
hand, during the HS the observed weak thermal component suggests
that the disk is truncated at a few hundreds/thousands gravitational
radii. The hard x-ray emission measured during this state is often
attributed to inverse Compton scattering of soft photons from the
outer disk by relativistic electrons in the hot inner region of the
system (Remillard \& McClintock 2006).
An example widely observed from radio to x-rays is the microquasar
GRS 1915+105
with a 10--18 solar mass black hole in the center of the binary
system (Mirabel \& Rodr\'{\i}guez 1994). Dhawan et al. (2000) have distinguished two main
radio states of this system, a plateau and a flare state. During the
plateau state the RXTE (2--12 keV) soft X-ray emission is weak,
while the BATSE (20--100 keV) hard X-rays are strong and the radio
flat spectrum is produced by a small scale nuclear jet. On the other
hand, during the flare phase, optically thin ejecta are
superluminally expelled up to thousands of AU and the radio spectral
index is between 0.5 and 0.8. The soft X-rays also flare during this
phase and exhibit high variability, while the hard X-rays fade for a
few days before recovering. In terms of the x-ray spectral states,
several works had verified that the radio flares of this source
occur during the SPLS  and the x-ray
emission of the plateau and the flare states would be different
manifestations of the SPLS (e.g, Reig et al. 2003).

The  radio and x-ray observations for other
microquasars (e.g, XTE J1859+226, XTE J1550-564) also suggest that
the ejection of relativistic matter happens when the source is very
active and in the SPLS (e.g, McClintock et. al 2007).
According to Fender et al. (2004), the origin of the optically thin
emission in radio is due to shock waves that are formed in the jet
when the system passes from a `hard' SPLS to a `soft' SPLS. However,
what generates these shock waves or the triggering mechanism for the
ejections of matter are not well defined.

de Gouveia Dal Pino \& Lazarian (2005) proposed that violent reconnection episodes between
the magnetic field lines of the inner disk region and those that are
anchored into the black hole
could  be responsible for the initial acceleration of the plasma jet to
relativistic speeds in GRS 1915+105. The same mechanism  could also lead to  the transition from the
`hard' SPLS to the `soft' SPLS seen in other microquasars (de Gouveia Dal Pino et al. 2010).
A detailed description of
the scenario adopted is given in de Gouveia Dal Pino \& Lazarian
(2005) and de Gouveia Dal Pino et al. (2010). Here we briefly present the main assumptions  made.
They have considered  a magnetized accretion disk around a black hole (BH) as schematized in Figure. 1. For simplicity, they assume that the inner zone of the accretion disk is
nearly co-rotating with the BH.
A magnetosphere around the central BH can be formed from the drag of
magnetic field lines by the accretion disk (e.g., MacDonald et al. 1986 and Wang et al. 2002). The disk coronal  poloidal magnetic field can be established
either by the action of a turbulent dynamo inside the accretion disk (King et al. 2004) or be advected  from outer regions. Recent simulations seem to show that the advection of external poloidal field can be effective (Beckwith, Hawley \& Krolik 2009). The advection of the magnetic flux with the gas towards the inner disk region will result in a gradual increase of the magnetic flux in this region (e.g., de Gouveia Dal Pino \& Lazarian 2005 and references therein; and McKinney \& Blandford 2009).

According to the magneto-centrifugal scenario (Blandford \& Payne 1982), this
poloidal magnetic flux summed to the disk differential rotation will
give rise to a wind that removes angular momentum from the system
increasing the accretion rate. This will increase the ram pressure
of the accreting material that will then push the magnetic lines in
the inner disk region towards the lines which are anchored in the BH
horizon, allowing a magnetic reconnection event to occur (see the Y-type zone labeled
as Helmet Streamer in Figure 1). Also, with the accumulation of the
poloidal flux in the inner regions the ratio between the
gas$+$radiation pressure to the magnetic field pressure ($\beta$)
will soon decrease. When the accretion
rate reaches values close to the Eddington limit and $\beta < 1$,
the magnetic reconnection event becomes violent and a large amount of magnetic energy
is released quickly by this process (de Gouveia Dal Pino et al. 2010).

To evaluate the
amount of magnetic energy that can be extracted through violent
reconnection, it is adopted
the standard  model (Shakura \& Syunyaev 1973) for the radiation-dominated accretion disk and Liu et al.'s model (Liu et al. 2002) to quantify the parameters
of the corona. Also, it is assumed  that the inner radius of the accretion disk
corresponds approximately to the last stable orbit around the BH
($R_X = 3 R_S$, where $R_S$ is the Schwarzschild radius), and  to determine the accretion rate immediately before an event of
violent magnetic reconnection, it is  assumed the equilibrium between
the disk gas ram pressure and the magnetic pressure of the magnetosphere
anchored at the BH horizon. It is  assumed further that  the intensity of the BH horizon  field is of the order of that of the inner disk (MacDonald et al. 1986). Under these conditions one can show that the
magnetic energy power released during violent magnetic reconnection is approximately given by (de Gouveia Dal Pino et al. 2010):

\begin{equation} \label{eq:MicroPot}
\dot{W}_{B} \cong 1,6 \times 10^{35} \alpha_{0.5}^{-19/16} \beta_{0.8}^{-9/16} M_{14}^{19/32} R_{X,7}^{-25/32} l_{100}^{11/16} \hspace{0.2cm} erg/s
\end{equation}

where the BH mass $M = 14
M_{\odot} M_{14}$ and  the disk inner radius  $R_{X} = 10^{7}$ cm $R_{X,7}$
are parameters suitable, for instance, for the microquasar GRS1915+105; $\alpha = 0.5\alpha_{0.5}$ is the disk viscosity; $\beta = 0.8 \beta_{0.8}$ is defined as the ratio between the  gas$+$radiation pressure (which is
dominated by radiation pressure) and the magnetic pressure; and $l_{100}= 100 R_X$ is the scale height of the Y neutral zone in the corona.
The corresponding reconnection time is
$t_{rec} \cong \frac{R_{X}}{\xi v_{A}} \cong 10^{-4} \xi^{-1} R_{X,7}$ s,
where $\xi = v_{rec}/v_A$  is the reconnection rate, $v_{rec}$ is
the reconnection velocity, and $v_A$ is the Alfv\'en speed. For the conditions around a BH we
find that $v_A \simeq c$. For fast reconnection $\xi$ can be as large as the observed values in the Sun, i.e.,  of a  few tenths (Kowal et al. 2009) and this relation indicates that the release of magnetic energy is also very fast, as required to produce flares.
Then energy stored in the magnetic
field will be released suddenly and at least part of it will be used
to accelerate charged particles. High speed particles
will spew outward giving rise to the relativistic radio ejecta and producing a luminous blob.

After reconnection, the destruction of the vertical magnetic flux in
the inner disk will make $\beta$ to increase and  the corona return to
a less magnetized condition with most of the energy being dissipated
locally in the disk, instead of in the outflow.
The enhanced x-ray emission that often accompanies the violent
flares in microquasars could be easily explained within the scenario
above  as due to the increase in the accretion rate
immediately before the violent magnetic reconnection events.
The observed soft x-ray emission is expected to be a fraction of the
accretion power and for an enhanced accretion rate near the
Eddington rate this is given by
$\dot{W}_{ac} \cong \frac{G M_{BN} \dot{M}}{R_{X}} \cong 1.87 \times 10^{39} M_{14} \dot{M}_{19} R_{X,7}^{-1}$ erg/s,
which is compatible with soft x-rays observations (Remillard \& McClintock 2006). On
the other hand, the hard x-ray component that is also often observed
could be explained by inverse Compton scattering of the soft x-rays
photons by the hot electrons of the corona/jet. If this is the case,
then we expect that after the radio flare the hard x-ray luminosity
will decrease because of the decrease of the number of
relativistic electrons (since most of them are accelerated away from
the system). In fact, this behavior is also compatible with the
observations (e.g., Dhawan et al 2000).

An  estimation of the electrum spectrum at the launching region produced in a magnetic reconnection  episode indicates that most of the relativistic electrons should be self-absorbed at this point.   When moving away, the produced plasmon cloud dilutes,
becoming transparent to its own radiation, first in the infrared and
then in radio frequency. The computation of the evolution of this
spectrum (e.g., Reynoso \& Romero 2009) is out of the scope of the present work as it requires the
building of a detailed model for the corona which will be considered
elsewhere.
Nonetheless, we can make some predictions by comparing
the calculated power released during a violent magnetic reconnection event with the
observed luminosities, as below.

Considering the same assumptions presented above, we obtain similar
scaling relations for AGNs. Figure 2 depicts a synthesis of the
magnetic reconnection scenario for
relativistic sources including both microquasars and AGNs. The
diagram shows the calculated magnetic power released in violent
magnetic reconnection events as a function of the central source
mass for a suitable choice of the parameter space. The symbols  correspond to the observed radio luminosities of superluminal
components (stars for microquasars, circles and triangles for the low luminous AGNs, i.e., LINERs (see below) and  Seyfert galaxies, respectively, and squares for luminous AGNs).

\begin{figure}
\begin{center}
 \includegraphics[width=0.47\textwidth]{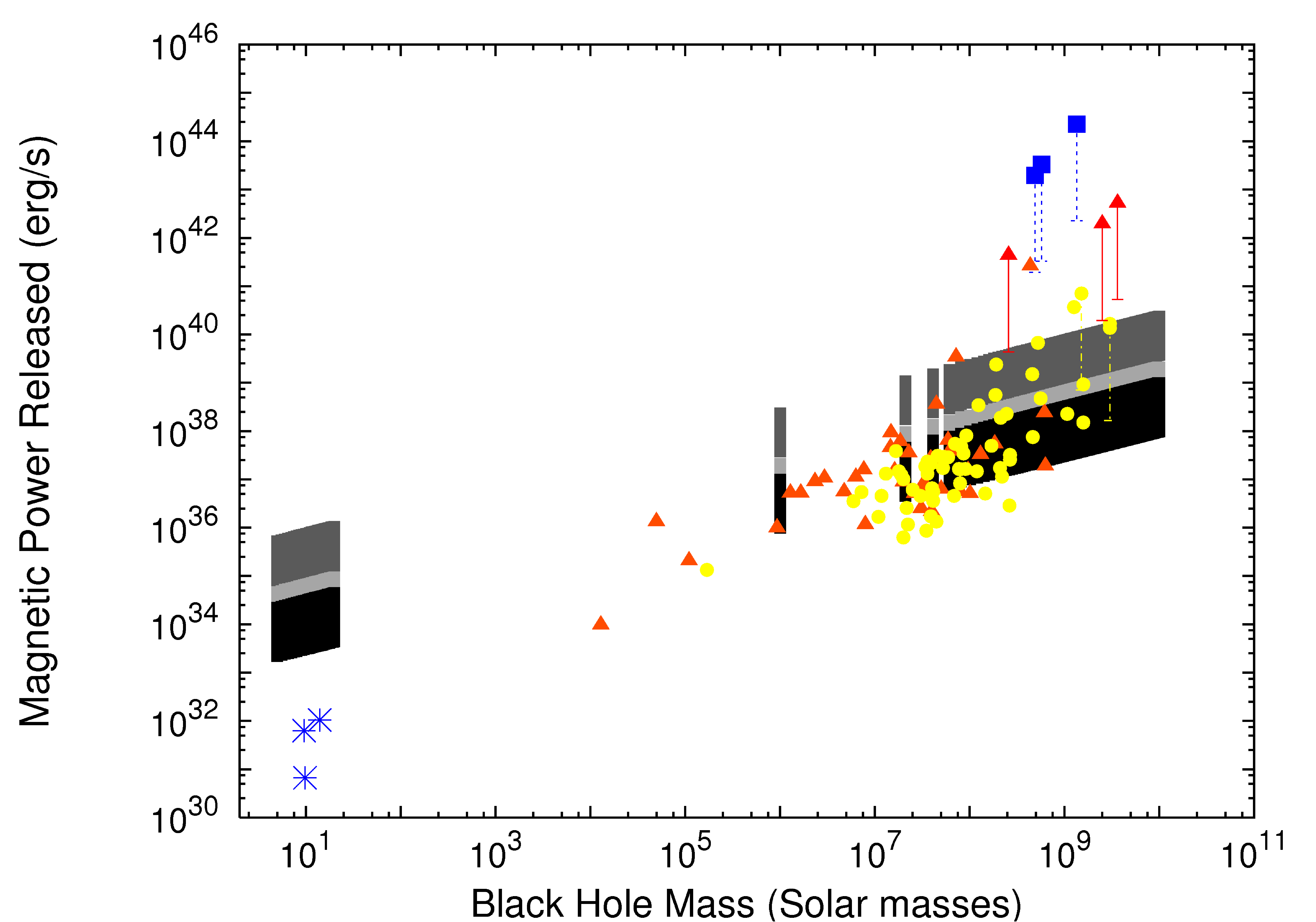}
\caption{$\dot{W}_B$ versus the BH mass $M$ for both microquasars and AGNs. The stars represent the observed radio luminosities for microquasars.  The circles, triangles and squares are observed radio luminosities of jets at parsec scales from LINERS,  Seyfert galaxies, and luminous AGNs (or quasars), respectively. The vertical thin bars associated to some sources stand for a reduction by a factor of 100 in the  (isotropic) observed luminosities due to  relativistic beaming. The thick bars correspond to the calculated magnetic reconnection power and encompass the parameter space that spans $5 M_{\odot} \leq M \leq 10^{10} M_{\odot}$, $0.05 \leq \alpha \leq 0.5$, $0.1 \leq \beta \leq 1$, and
$1 R_S \leq l \leq 1000 R_S$ (or $0.3 R_X \leq l \leq 333 R_X$), with $1 R_S \leq l \leq 10 R_S$ in black, $10 R_S < l \leq 30 R_S$ in light gray, and $30 R_S < l \leq 1000 R_S$ in dark gray. (Extracted from de Gouveia Dal Pino et al. (2010)}
\end{center}
\end{figure}

Figure 2 indicates that the magnetic power released during violent reconnection events is able to explain the emission of relativistic radio blobs from both microquasars and the so called low-luminous AGNs (LLAGNs) which include  Seyfert galaxies and LINERs (these later have emission-line luminosities that are typically a factor of $10^2$ orders of magnitude smaller than those of luminous AGNs; e.g., Ho et al. 1997). This establishes a correlation between the magnetic reconnection power of stellar mass and supermassive black holes according to Eq. (2)  spanning over $10^9$ orders of magnitude in  mass of the sources.


de Gouveia Dal Pino \& Lazarian (2005) have proposed a mechanism to
accelerate particles to relativistic velocities within the
reconnection zone,  in a  similar process to the first-order Fermi.
Charged particles are confined in the reconnection zone by the
particle+magnetic flux coming from both sides of the current sheet
in such a way that their energies increase stochastically. They  have shown that a power-law
electron distribution with $N(E) \propto E^{-5/2}$ and a synchrotron
radio power-law spectrum $S_{\nu} \propto \nu^{-0.75}$ can be
produced in this case
\footnote{This mechanism does not remove the possibility that further out the
relativistic fluid may be also produced behind shocks which are
formed by the magnetic plasmons that erupt from the reconnection
zone. Behind these shocks a standard first-order Fermi acceleration
may also occur resulting a particle power-law spectrum $N(E) \propto
E^{-2}$ and a synchrotron spectrum $S_{\nu} \propto \nu^{-0.5}$.
Both radio spectral indices are consistent with the observed
spectral range during the flares, e.g.,  of GRS 1915-105 ($-0.2 < \alpha_{R} <
-1.0$; Dhawan et al. 2000).}
Kowal, de Gouveia Dal Pino \& Lazarian (2010) are
presently testing this acceleration model in reconnection sites
using 3D Godunov-MHD simulations combined with a particle in-cell
technique. The diagrams in Figure 3 show preliminary tests that compare the energy increase of the particles as a function of time due to acceleration in a pure MHD turbulent medium, where instantaneous reconnection can occur between turbulent structures with opposing magnetic field polarity, and acceleration inside a current sheet where $fast$ $reconnection$ takes place as described in section 2. We see that after an initial slower increase, the energy rate increase is much larger in the environment with fast reconnection.

\begin{figure}
\begin{center}
 \includegraphics[width=0.47\textwidth]{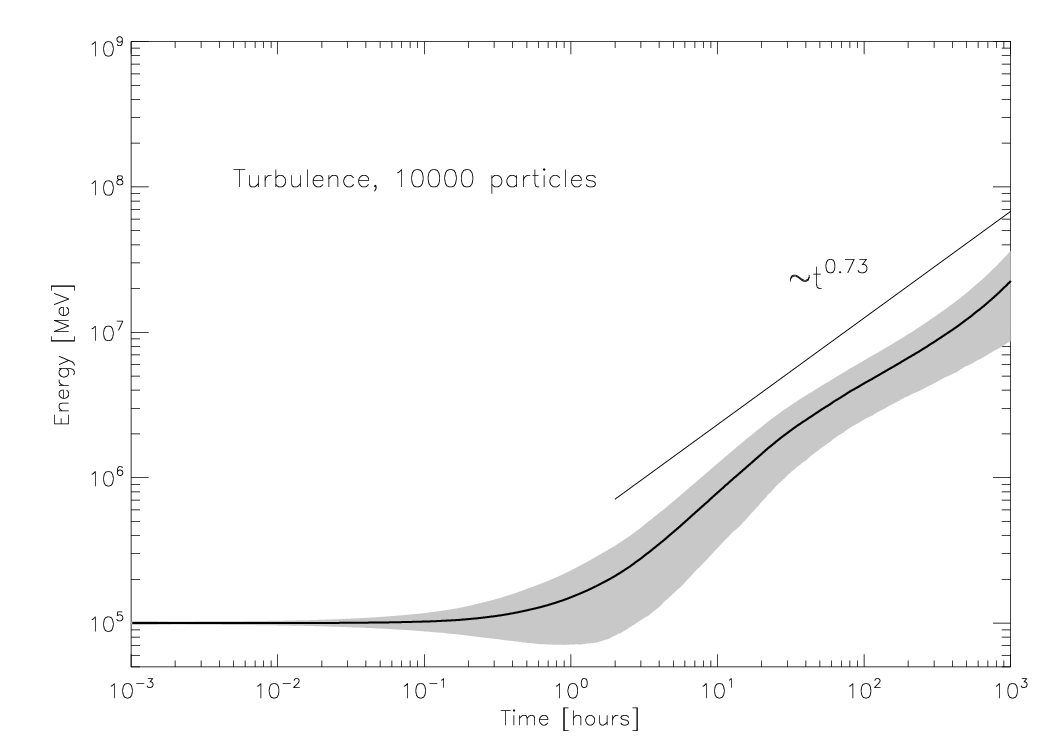}
 \includegraphics[width=0.47\textwidth]{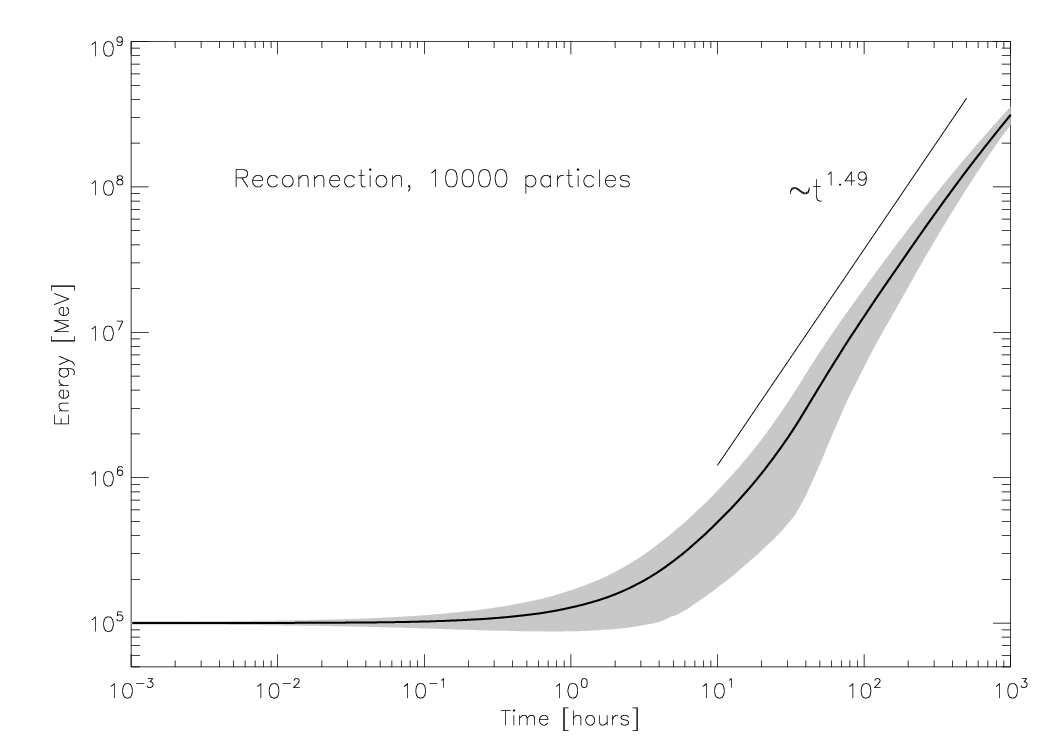}
\caption{Comparison of particle acceleration in a pure MHD turbulent medium (left) and a turbulent medium inside a current sheet where fast reconnection takes place (right) (Kowal, de Gouveia Dal Pino \& Lazarian, (2010)). }
\end{center}
\end{figure}

\section{Summary and Conclusions}
In this paper we reviewed  the model proposed by de Gouveia Dal Pino
\& Lazarian (2005) and de Gouveia Dal Pino et al. (2010) for violent reconnection episodes occurring at jet launching basis between the magnetic field lines of
the inner accretion disk region and those that are anchored into the central source.

In the case of relativistic jets this may heat the coronal/disk gas and accelerate the plasma to relativistic velocities through a diffusive first-order Fermi-like process within the reconnection site which   can produce intermittent relativistic ejections or plasmons. The resulting power-law electron distribution is compatible with the synchrotron radio spectrum observed during the outbursts of these sources. Preliminary numerical tests indicate that this acceleration mechanism within the reconnection site is promising. A diagram of the magnetic energy rate released by violent reconnection as a function of the black hole (BH) mass spanning $10^9$ orders of magnitude shows that the estimated magnetic reconnection power is sufficient to explain the observed radio luminosities of the outbursts, from microquasars to low luminous AGNs. This correlation does not hold for radio loud AGNS. These  would require reconnection events with super-Eddington accretion in order to explain the formation and  emission of relativistic blobs as due to reconnection. This is possibly because their surroundings are much denser and then "mask" the emission due to  coronal magnetic activity at sub-Eddington rates. In this case, particle re-acceleration behind shocks will probably prevail further out in the jet launching region and will be responsible for the radio emission.

In addition, the magnetic reconnection events cause the heating of the coronal gas which can be conducted back to the disk to enhance its thermal soft x-ray emission as observed during outbursts in microquasars. The decay of the hard x-ray emission after a radio flare could also be explained in this model due to the escape of relativistic electrons (which could produce hard x-ray by inverse Compton scattering of soft x-rays) with the evolving jet outburst.

In the case of YSOs, a similar magnetic reconnection model could produce the observed x-ray flares in some sources and heat the jet launching base.

This work was partially supported by grants of the Brazilian Agencies FAPESP and CNPq and by a grant of the MPIA- Garching.



\begin{thebibliography}{0}    


\bibitem{blandfordpayne} Blandford, R.~D., \& Payne, D.~G., 1982, MNRAS, 199, 883
\bibitem{McKinney Blandford 2009} McKinney, J.~C., \& Blandford, R.~D., 2009, MNRAS, 394, 126
\bibitem{de Gouveia Dal Pino 2005} de Gouveia Dal Pino, E.~M., 2005, Advances in Space Research, 35, 908
\bibitem{Balbus & Hawley 1998} Balbus, S.~A., \& Hawley J.~F., 1998, RvMP, 70, 1
\bibitem{de Gouveia Dal Pino & Lazarian 2005} de Gouveia Dal Pino, E.~M., \& Lazarian, A., 2005, A\&A, 441, 845
\bibitem{dGDP et al. 2010} de Gouveia Dal Pino, E. M., Piovezan, P., Kadowaki, L. H. S., 2010, A\&A (submitted)
\bibitem{L&V 1999} Lazarian, A., \& Vishniac, E., 1999, ApJ, 517, 700
\bibitem{Kowal et al. 2009} Kowal, de Gouveia Dal Pino \& Lazarian 2009, in prep.
\bibitem{Bacciotti et al. 2002} Bacciotti, F., Ray, T.~P., Mundt, R., Eislöffel, \& J., Solf, J., 2002, ApJ, 576, 222
\bibitem{Cerqueira et al. 2007} Cerqueira, A. H., Velázquez, P. F., Raga, A. C., Vasconcelos, M. J., de Colle, F., 2006 A\&A,  448, 231
\bibitem{Soker 2009} Soker, N. in {\it Astrophysical Outflows and Associated Accretion Phenomena}, eds. E. M. de Gouveia Dal Pino, A. Raga, IUA JD 7 at the XXVIIth IAU General Assembly, {\it IAU Highlights of Astronomy}, 15, eds. I. F Corbett et al., 2009 (in press), 2009arXiv0909.4847S
\bibitem{Alencar_Johns-Krull & Basri 2001} Alencar, S. H. P.; Johns-Krull, C. M.; Basri, G., 2001, ApJ, 122, 3335
\bibitem{Camenzind 1990} Camenzind, M., 1990,  Rev. Modern Astron.,  3,  234 
\bibitem{Goodson & Winglee 1999} Goodson, A.P. \& Winglee, R.M. 1999, ApJ, 524, 159
\bibitem{Romanova_Ustyugova_Koldoba et al. 2004} Romanova, M.M., Ustyugova, G.V., Koldoba, A.V., \&  Lovelace, R.V.E. 2004, ApJ, 616, L151
\bibitem{Alencar 2007} Alencar, S. H. P., 2007, in Star-Disk Interaction in Young Stars, Proceedings of the IAU Symposium, 243, 71
\bibitem{Bouvier et al. 2007} Bouvier, J., Alencar, S. H. P., Boutelier, T., et al., 2007, A\&A, 463, 1017
\bibitem{Feigelson & Montmerle 1999} Feigelson, E.~D., \& Montmerle, T., 1999, ARA\&A, 37, 363
\bibitem{Favata et al. 2005} Favata, F., Flaccomio, E., Reale, F., et al., 2005, ApJs, 160, 469
\bibitem{Sakura & Syunyaev 1973} Shakura, N.~I., \& Syunyaev, R.~A., 1973, A\&A, 24, 337
\bibitem{Ray07} Ray, T., Dougados, C., Bacciotti, F., Eisl{\"o}ffel, J., \& Chrysostomou, A., 2007, Protostars and Planets V, 231
\bibitem{Cerqueira & de Gouveia dal Pino 2004} Cerqueira, A.~H., \& de Gouveia Dal Pino, E.~M., 2004, A\&A, 426, L25
\bibitem{Remillard & McClintock 2006} Remillard, R.~A., \& McClintock, J.~E., 2006, ARA\&A, 44, 49
\bibitem{Mirabel & Rodriguez 1994} Mirabel, I.~F., \& Rodr{\'{\i}}guez, L.~F., 1994, Nature, 371, 46
\bibitem{Dhawan et al. 2000} Dhawan, V., Mirabel, I.~F., \& Rodr{\'{\i}}guez, L.~F., 2000, ApJ, 543, 3735
\bibitem{Reig et al. 2003} Reig, P., Belloni, T., \& van der Klis, M., 2003, A\&A, 412, 229
\bibitem{McClintock et. al 2007} McClintock, J.~E., Remillard, R.~A., Rupen, M.~P., et al., 2007, ArXiv e-prints, 705
\bibitem{Fender et al. 2004}Fender, R.~P., Belloni, T.~M., \& Gallo, E., 2004, MNRAS, 355, 1105
\bibitem{Macdonald et al. 1986} MacDonald, D.~A., Thorne, K.~S., Price, R.~H., \& Zhang, X.~H., 1986, in Black Holes: The Membrane Paradigm, 120
\bibitem{Wang et al. 2002} Wang, D.~X., Xiao, K., \& Lei, W.~H., 2002, MNRAS, 335, 65
\bibitem{King et al. 2004} King, A.~R., Pringle, J.~E., West, R.~G., \& Livio, M., 2004, MNRAS, 348, 111
\bibitem{Beckwith_Hawley & Krolik 2009} Beckwith, K., Hawley, J.~F, \& Krolik, J.~H., 2009, ApJ, 707, 428
\bibitem{McKinney & Blandford 2009} McKinney, J.~C., \& Blandford, R.~D., 2009, MNRAS, 394, 126
\bibitem{Liu et al. 2002} Liu, B.~F., Mineshige, S., \& Shibata, K., 2002, ApJ, 572, L173
\bibitem{Reynoso & Romero 2009}Reynoso, M. M.; \& Romero, G. E., 2009, A\&A, 493, 1
\bibitem{Ho et al. 1997} Ho, L.~C., Filippenko, A.~V., \& Sargent, W.~L.~W., 1997b, ApJS, 112, 315
\bibitem{Kowal_de Gouveia Dal Pino & Lazarian 2010} Kowal, G., de Gouveia Dal Pino, E. M.,  \& Lazarian, A. 2010 (in prep.)










\end{thebibliography}
\end{document}